\documentstyle[12pt,epsfig]{article}
        
\def\beq{\begin{equation}}
\def\eeq#1{\label{#1}\end{equation}}
\def\eeqn{\end{equation}}
\def\beqa{\begin{eqnarray}}
\def\eeqa#1{\label{#1}\end{eqnarray}}
\def\eeqan{\end{eqnarray}}
\def\CR{\nonumber \\ }
\def\leqn#1{(\ref{#1})}

\def\({\left(}
\def\){\right)}
\def\B{$B_H$}

\def\to{\rightarrow}

\def\arX#1{[arXiv:hep-ph/#1]}
\def\arXiv#1{arXiv:hep-ph/#1}
\def\astr#1{[arXiv:astro-ph/#1]}
\def\astro#1{arXiv:astro-ph/#1}

\newcommand{\text}[1]{\rm #1}

\def\twogam{$\gamma\gamma$}
\def\Zgamma{$Z\gamma$}

\def\stacksymbols #1#2#3#4{\def\theguybelow{#2}
    \def\vp{\lower#3pt}
    \def\sp{\baselineskip0pt\lineskip#4pt}
    \mathrel{\mathpalette\intermediary#1}}

\def\intermediary#1#2{\vp\vbox{\sp
     \everycr={}\tabskip0pt
     \halign{$\mathsurround0pt#1\hfil##\hfil$\crcr#2\crcr
              \theguybelow\crcr}}}

\def\gapproxeq{\stacksymbols{>}{\sim}{2.5}{.2}}

\begin{document}

\begin{titlepage}
\begin{flushright}
{\tt hep-ph/0610357} \\
\end{flushright}

\vskip.5cm
\begin{center}
{\huge \bf Indirect Detection of Little Higgs \\ Dark Matter} \vskip.2cm
\end{center}
\vskip1cm

\begin{center}
{\bf \large Maxim Perelstein and Andrew Spray} 
\end{center}
\vskip 8pt

\begin{center}
{\it Cornell Institute for High-Energy Phenomenology, Cornell University, 
Ithaca, NY~14853}
\end{center}

\vglue 0.3truecm

\begin{abstract}
\vskip 3pt \noindent
Little Higgs models with T parity contain an attractive dark matter
candidate, the heavy photon. We compute the cross section of the heavy photon 
annihilation into Z-photon pairs, which turns out to be substantially 
higher than the previously computed cross section for the two photon final 
state. Unfortunately, even with this enhancement, the monochromatic photon 
flux from galactic heavy photon annihilation is unlikely to be detectable by 
GLAST or the currently operating atmospheric Cerenkov telescopes. 
We also compute the flux of high-energy neutrinos from the annihilation
of the heavy photons captured by the Sun and the Earth. The maximum flux of 
upward-going muons due to such neutrinos is about 1~yr$^{-1}$km$^{-2}$.
\end{abstract}

\end{titlepage}

\section{Introduction} 

The presence of dark matter in the universe has been firmly established by 
observations of galaxy rotation curves, large scale structure, and cosmic 
microwave background radiation. The microscopic nature of dark matter, 
however, remains unknown. 
One attractive scenario is that it consists of stable weakly-interacting 
massive particles (WIMPs), with masses around the weak scale, $\sim$100-1000 
GeV. If the WIMPs were in thermal equilibrium with other species 
in the early universe, their relic abundance naturally matches the 
observed dark matter density. Further, many theoretical extensions of 
the Standard Model (SM) of particle physics include such particles. The most 
famous example is no doubt the neutralino of most SuperSymmetric (SUSY) 
models; another well known example is the lightest Kaluza-Klein excitation in 
models with universal extra dimensions.

In this paper, we consider the dark matter candidate emerging in another 
popular class of theories, the Little Higgs models~\cite{LH} (for reviews, 
see~\cite{ST1,P1}). In these theories, the Higgs is a pseudo-Nambu-Goldstone 
boson associated with spontaneous global symmetry breaking in an extended 
electroweak sector, which occurs at a scale $f\sim1$ TeV. The Higgs acquires
a mass due to an explicit breaking of the global symmetries by gauge and
Yukawa interactions, but a special ``collective'' manner in which this
explicit breaking occurs ensures that the Higgs is relatively light, 
$m_h\ll f$. Several models based on this idea 
have been constructed, most of which are tightly constrained by precision
electroweak data. However, introducing an additional discrete symmetry, the
T parity, weakens the precision electroweak constraints 
significantly~\cite{CL}. The Littlest Higgs model with T parity 
(LHT)~\cite{LHT} provides an example of a fully realistic and natural model 
consistent with precision electroweak data~\cite{HMNP}, and is sufficiently 
simple to allow for detailed phenomenological 
analyses~\cite{HM,Wyler,MSU,CHPV}.

An interesting consequence of T parity is the stability of the lightest T-odd 
particle (LTP), typically the heavy partner of the hypercharge gauge boson 
$B_H$. (This particle is often referred to as the ``heavy photon'', even 
though it does {\it not} couple to the electric charge.) The \B~is 
weakly interacting, has a mass in the range $80-500$ GeV, and it was 
shown to have the correct relic abundance to account for all of the 
observed dark matter in certain 
regions of the parameter space~\cite{HM,BPNS}. Potential non-gravitational
signatures of the heavy photon dark matter have been  discussed in~\cite{BPNS} 
(direct detection and indirect detection via high-energy photons 
produced by \B~annihilations in the galactic center) and 
in~\cite{Okada} (indirect detection via high-energy positrons). In this
paper, we continue the evaluation of indirect signatures for this
dark matter candidate. First, we complete the previous analysis of the 
monochromatic photon 
flux~\cite{BPNS} by computing the flux from the $B_H B_H\to Z\gamma$ 
reaction, whose cross section turns out to be substantially larger than the
$B_H B_H\to 2\gamma$ reaction analyzed in~\cite{BPNS}. Unfortunately,
even with this enhancement the near-future prospects for the detection of the 
gamma line do not look promising, especially when the EGRET constraint on the
continuous photon flux is taken into account. Second, we evaluate the 
neutrino flux from the annihilation of the heavy photons captured by the Sun 
and the Earth. While the predicted fluxes are too low to be observed by
the near-future detectors such as ICECUBE, a further improvement of 
two-three orders of magnitude in detector sensitivity would allow to probe 
interesting parts of the parameter space.

\section{The Model}

The LHT model has been described in detail elsewhere~\cite{LHT,HMNP,HM}; 
here, we briefly 
summarize the features of the model important for our analysis. The  
extended electroweak sector of the LHT model has a global $SU(5)$ symmetry,
which is spontaneously broken to $SO(5)$ at a scale $f\sim 1$ TeV.
At energies below the cutoff $\Lambda\sim 4\pi f$, the dynamics of this sector 
is described by a non-linear sigma model. To incorporate gauge interactions,
the subgroup ${\cal G}=(SU(2)\times U(1))^2$ of the $SU(5)$ is weakly gauged. 
The T partity interchanges the two $SU(2)\times U(1)$ factors. At the
scale $f$, the gauge symmetry ${\cal G}$ is broken down to $SU(2)\times U(1)$,
identified with the SM electroweak group. The gauge bosons corresponding to
the broken generators, $W^\pm_H$, $W^0_H$ and \B, are T-odd and
acquire masses at the scale $f$:\footnote{In Eq.~\leqn{masses} and throughout 
this paper, we neglect corrections of order $v^2/f^2$.}
\beq
M(W_H)\approx gf\,,\qquad M\equiv M(B_H)\approx\frac{g^\prime f}{\sqrt{5}}
\approx 0.16 f.
\eeq{masses}
The uneaten $SU(5)/SO(5)$ Nambu-Goldstone bosons decompose into a T-even 
$SU(2)$ doublet $H$, identified with the SM Higgs, and a T-odd $SU(2)$ 
triplet $\Phi$, which acquires a mass $m_\phi=\sqrt{2}m_hf/v$ at one loop. 
In the fermion sector, the LHT model contains vector-like T-odd partners 
for the left-handed SM quarks and leptons. We will assume a common mass
scale $\tilde{M}$ for all of these particles. This choice is motivated both
by simplicity and (in the case of the T-odd quarks) by flavor 
constraints~\cite{flavor}. Finally, the model contains an additional pair
of weak-singlet, charge-2/3 quarks: $T_+$ (T-even) and $T_-$ (T-odd). 
These are required to ensure the cancellation of the one-loop quadratic 
divergence in the Higgs mass parameter from the SM top. The mass $M_{T-}$ is 
a free parameter, and
\beq
M_{T_+}=M_{T_-}\left( 1-\frac{m_t^2f^2}{v^2M_{T_-}^2}\right)^{-1/2}.
\eeq{newtms}
With our assumptions, the spectrum and the (renormalizable) couplings of
the LHT model are completely described in terms of the three parameters, 
$f$, $\tilde{M}$, and $M_{T_-}$, in addition to the familiar parameters of 
the SM. 

Due to the smallness of $g^\prime$ and the favorable group theory factor, 
the ``heavy photon'' $B_H$ is substantially lighter than $f$, and is always 
the lightest among the T-odd bosons of the theory. The fermion mass 
parameters $\tilde{M}$ and $M_{T_-}$ are at the scale $f$, so it is 
natural to consider the part of the parameter space where the $B_H$ is the 
lighest T-odd particle (LTP). In this region, the heavy photon is stable and 
can contribute to dark matter. In this paper we assume that all of the 
observed dark matter density is due to the \B. Calculations of the relic 
density have shown~\cite{HM,BPNS} that there are three regions of the 
parameter space in which this is possible.  In the first two, the \B~density 
in the early universe is controlled by \B~pair-annihilation through an 
$s$-channel Higgs.  The correct relic density is obtained when the 
center of mass energy of a \B~\B~ collision in the nonrelativistic regime
is near, but slightly displaced from, the resonance. There are typically
two solutions, one on either side of the resonance.
In the third region, the \B~relic density is set by 
coannihilation processes with T-odd fermions, and hence $M$ must be close to 
$\tilde{M}$. The precise measurement of the present dark matter density 
implies tight correlations between the model parameters in these regions.
It was found in~\cite{BPNS} that they can be approximately described by
\beq
M=\left\{\begin{array}{ll}(m_h/2.38)-10&\textrm{``Low'' pair annihilation 
region}\\ 
(m_h/1.89)+44&\textrm{``High'' pair annihilation region}\\ 
\tilde{M}-20&\textrm{coannihilation region}\end{array}\right.
\eeq{regions}
(All masses are in GeV.)  With the reduced uncertainties from the 3-year
WMAP data set~\cite{WMAP3}, $\Omega_{\rm dm}h^2 = 0.104\pm 0.009$, the allowed 
variation of $M$ around the central values in~\leqn{regions} is at most 
of order $\pm 5$ GeV at the 2$\sigma$ level. We will thus treat the allowed 
regions as lines defined by Eq.~\leqn{regions}.

\section{Monochromatic Photon Flux from \B\B$\to Z\gamma/h\gamma$ 
Annihilation}

A promising avenue for indirect WIMP detection is through high energy gamma 
rays~\cite{BUB1}. Since the galaxy is transparent to photons in the
interesting energy range, any features in the photon spectrum are preserved. 
This can both aid in distinguishing the signal from the background, and 
provide 
information about the WIMP properties. In particular, a pair-annihilation of 
WIMPs into two-body final states containing photons leads to a flux
of nearly monochromatic gamma rays, allowing for efficient background 
subtraction. In the context of the LHT model, the monochromatic gamma ray 
flux from the process \B\B$\to 2\gamma$ was analyzed in Ref.~\cite{BPNS}.
In this section, we will analyze the additional two processes which
produce monochromatic photons, \B\B$\to Z\gamma$~or $h\gamma$.

The \B\B~annihiliation in the LHT model is dominated by processes with 
an $s$-channel Higgs boson exchange.  
The cross section for producing a given final state $X$ in this channel 
can be related to the partial decay width of an (off-shell) Higgs~\cite{BPNS}:
\beq
\sigma_X u\equiv\sigma\left(B_HB_H\to X\right)u=\frac{g'^4v^2}{72M^4}
\frac{s^2-4sM^2+12M^4}{\left(s-m_h^2\right)^2+m_h^2\Gamma_h^2}
\frac{\hat{\Gamma}\left(h\to X\right)}{\sqrt{s}}.
\eeq{xsecxp}
Here, $u$ is the relative velocity of the annihilating WIMPs; $s\approx 4M^2$ 
in the galactic centre; and the hat on $\Gamma$ indicates that the 
substitution $m_h\to\sqrt{s}$ should be made in the expression for the
on-shell Higgs decay width. 

Using the well-known results for the partial Higgs decay width in the 
$Z\gamma$ channel~\cite{SVVZ1,THHG1}, we obtain\footnote{We neglect the 
contribution to the Higgs width from the loops of the T-odd scalar $\Phi$.}
\beq
\hat{\Gamma}\left(h\to Z\gamma\right)=\frac{\alpha\,g^2}{2048\pi^4}
\frac{s^{3/2}}{m_W^2}\mid{\mathcal A}_F+{\mathcal A}_G\mid^2
\left(1-\frac{m_Z^2}{s}\right)^3.
\eeq{ratezp}
Here, $\mathcal{A}_F$ and $\mathcal{A}_G$ are the contributions to the 
matrix element from fermions and gauge bosons, respectively:
\beqa
{\mathcal A}_F &=& -4\sum_f \frac{\sqrt{2}y_f\tau_W^{1/2}}{g\tau_f^{1/2}}
N_{c_f}Q_fV_f\left[I_1\left(\tau_f,\lambda_f\right)
-I_2\left(\tau_f,\lambda_f\right)\right]\,,\CR
\mathcal{A}_G &=& \hskip-2mm -\sum_g \frac{c_g \tau_W}{\tau_g}
V_gQ_g\left\{4(3-t_W^2)I_2(\tau_g,\lambda_g)
+\left[\left(1+\frac{2}{\tau_g}\right)t_W^2-\left(5+\frac{2}{\tau_g}
\right)\right]I_1(\tau_g,\lambda_g)\right\}\,,
\eeqa{zpfun}
where $t_W=0.548$ is the tangent of the Weinberg angle, and
\beq
\tau_i\equiv\frac{4m_i^2}{s}\,,\qquad\lambda_i\equiv\frac{4m_i^2}{m_Z^2}\,.
\eeq{mafrdf}
The sums run over all particles of the relevant type (including both the 
SM and the additional particles of the LHT model), with the electric charge,
the multiplicity and the coupling to the $Z$ boson of each particle given by 
$Q_i$, $N_{c_i}$, and $V_i$, respectively.  
For fermions, $V_i$ denotes the vector part of
the $i\bar{i}Z$ coupling; there is no contribution from the axial part.
In particular, for the extra vector-like fermions of the LHT model, we obtain
\beq
V_i=\frac{g}{c_W}\left(T_{3}(i)-s_W^2 Q(i)\right)\,,
\eeq{Vextra}
where $T_3(\tilde{U}, \tilde{N})=1/2,~T_3(\tilde{D}, 
\tilde{E})=-1/2$, and $T_3(T_{+},T_{-})=0$. The fermion and vector boson
trilinear couplings to the Higgs are given by $y_f/\sqrt{2}$ and
$c_i g M_W \eta^{\mu\nu}$, respectively. (With this normalization, $y_f$ are
the standard Yukawa couplings for the SM fermions and $c_W=1$ for the SM 
$W^\pm$ boson.) The functions $I_{1,2}$ are given by 
\beqa
I_1(a,b)&=&\frac{a\,b}{2(a-b)}+\frac{a^2b^2}{2(a-b)^2}\left[f(a)-f(b)\right]+
\frac{a^2b}{(a-b)^2}\left[g(a)-g(b)\right]\,,\CR
I_2(a,b)&=&-\frac{a\,b}{2(a-b)}\left[f(a)-f(b)\right]\,,
\eeqa{I}
where
\beq
f(x)=\left\{\begin{array}{ll}\left[\sin^{-1}\left(\sqrt{\frac{1}{x}}
\right)\right]^2&\textrm{if}\,x>1\\-\frac{1}{4}\left[\log\left(\frac{1+
\sqrt{1-x}}{1-\sqrt{1-x}}\right)-i\pi\right]^2&\textrm{if}\,x<1\end{array}
\right. 
\eeq{intfunf}
\beq
g(x)=\left\{\begin{array}{ll}\sqrt{x-1}\sin^{-1}\left(\sqrt{\frac{1}{x}}
\right)&\textrm{if}\,x>1\\ \frac{1}{2}\sqrt{1-x}\left[\log\left(\frac{1+
\sqrt{1-x}}{1-\sqrt{1-x}}\right)-i\pi\right]&\textrm{if}\,x<1\end{array}
\right.
\eeq{intfung}

For a telescope with line of sight parameterized by $\Psi=(\theta ,\phi)$ and 
an angular acceptance $\Delta\Omega$, the anomalous photon flux due to
\B\B$\to Z\gamma$ is given by~\cite{BUB1}
\beq
\Phi=\left(5.5\times 10^{-10}\,~{\rm s}^{-1}\,~{\rm cm}^{-2}\right)\left(\frac{
\sigma_{Z\gamma}u}{1\,\textrm{pb}}\right)\left(\frac{100\,\textrm{GeV}}{M}
\right)^2\bar{J}\left(\Psi ,\Delta\Omega\right)\Delta\Omega,
\eeq{fluxzp}
where the function $\bar{J}$ contains all information about the dark matter  
distribution in the halo. The photon energy is 
\beq
E_\gamma = M\left( 1 - \frac{M_Z^2}{4M^2}\right)\,.
\eeq{egamma} 
The thermal broadening of the line is much smaller than the energy resolution 
of any existing telescope, and can be neglected.

\begin{figure}
\begin{center}
\includegraphics[width=10cm]{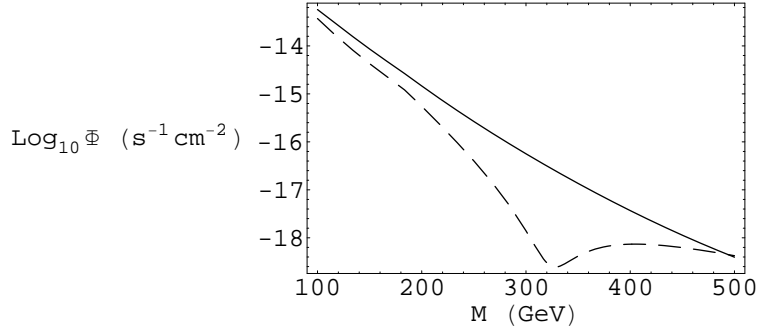}
\vskip-12mm
\caption{Photon fluxes from WIMP annihilation into \Zgamma~(solid line) and 
\twogam~(dashed line), in the ``high'' pair-annihilation region of the LHT 
model. The fluxes in the ``low'' pair-annihilation and coannihilation 
regions 
are similar. The fluxes are normalized to $\bar{J}\Delta\Omega=1$, and scale
linearly with this parameter.}
\label{Zgflux}
\end{center}
\end{figure}

The flux expected in the ``high'' pair-annihilation 
region of the LHT model\footnote{In all calculations of photon and neutrino 
fluxes, we assume $\tilde{M}=M_{T_-}=f$ in the pair-annihilation regions
and $M_{T_-}=f$ in the coannihilation region. The impact of varying these 
parameters on the flux predictions
is very small.} is plotted in Fig.~\ref{Zgflux}. For
comparison, the flux due to \B\B$\to2\gamma$, computed in~\cite{BPNS}, 
is also shown. Throughout the parameter space, the $Z\gamma$
final state provides a stronger monochromatic photon flux, with the ratio
of the $Z\gamma$ to $\gamma\gamma$ flux varying between about $1.5$ and 100.
(The largest values of this ratio are obtained for $M \sim 300$ GeV,
where the $\gamma\gamma$ cross section is supressed due to an 
accidental cancellation.) 
Similar results are obtained in the ``low'' pair-annihilation and the 
coannihilation regions of the parameter space.

Experimental searches for the anomalous high-energy gamma rays 
using Atmospheric Cerenkov telescopes (ACTs), such as HESS~\cite{HESS} 
and VERITAS~\cite{VERITAS}, are currently under way, and the space-based 
GLAST telescope~\cite{GLAST} is expected to perform such a search starting 
in 2007. Numerical simulations indicate that dark matter density may have
a sharp peak at the galactic center, in which case the flux of the
anomalous gamma rays would be maximized for a line of sight towards the
center of the Milky Way. The principal source of background for a search 
focusing on the galactic center region is the recently discovered powerful 
point-like gamma ray source~\cite{K1}, whose spectrum strongly suggests 
that its nature is
astrophysical. Taking into account this background, Zaharijas and 
Hooper~\cite{ZH} estimate that the minimal WIMP-related monochromatic 
photon flux 
required for 5$\sigma$ discovery is about $10^{-11}$ cm$^{-2}$sec$^{-1}$ 
for GLAST (which is sensitive to photon energies up to about 300 GeV) and 
$10^{-12}$ cm$^{-2}$sec$^{-1}$ for the ACTs (sensitive to $E_\gamma 
\gapproxeq 200$ GeV). The $Z\gamma$ line predicted in the LHT model would be
observable at GLAST and the ACTs only if the fluxes are enhanced by a 
strong spike in the dark matter concentration around the galactic center: 
depending on the \B~mass, values of $\bar{J}\Delta\Omega$ in the $10^2
\ldots 10^4$ range are required. Many (though not all) models of the galactic 
halo contain such spikes: for example, the profile of Moore et 
al.~\cite{Moore} predicts  $\bar{J}\approx10^5$ for $\Delta\Omega=10^{-3}$ sr.

\begin{figure}
\begin{center}
\includegraphics[width=10cm]{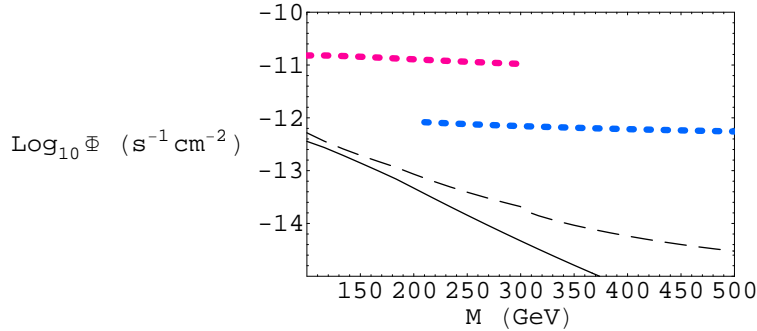}
\vskip-12mm
\caption{Maximum $Z\gamma$ photon fluxes consistent with the EGRET bound on
the continuous photon flux from WIMP annihilation. Solid line:
``high'' pair-annihilation region; dashed line: coannihilation region with
$m_h=300$ GeV. Also shown are the sensitivities of GLAST (dotted red
line) and HESS (dotted blue line)~\cite{ZH}.}
\label{fig:maxflux}
\end{center}
\end{figure}

However, the prospects of the future searches are further restricted by
the constraints on the {\it continuous} component of the photon flux from 
the observations of the gamma rays from the galactic center in the energy 
range up to 30 GeV by EGRET~\cite{EGRETcon,ZH}. The continuous component 
of the flux in the LHT model was computed in Ref.~\cite{BPNS}. The 
{\it maximum} value of the $Z\gamma$ photon flux compatible with the
EGRET constraint (independently of the galactic halo profile) is shown in 
Fig.~\ref{fig:maxflux}. It is clear that the maximum flux is substantially 
below the sensitivity of the current and near-future telescopes throughout
the parameter space. 

Given the presence of a point-like background source at the galactic center, 
other regions with possible dark matter overdensity were suggested as
potential targets for a search for anomalous gamma rays. These include
dwarf spheroidal companion galaxies to the Milky Way such as Saggitarius,
Draco and Canis Major~\cite{EFS1,dwarf}, as well as the Large Magellanic Cloud
and the M87 galaxy~\cite{NPS1}. In addition, in models where the galaxy is 
built up from hierarchical dark matter clustering one should expect localized 
clumps of dark matter inside the Milky Way halo. The values of 
$\bar{J}\Delta\Omega$ expected for these objects are model-dependent. For 
example, the dark matter profiles in dwarf spheroidals surveyed in 
Ref.~\cite{EFS1} give $\bar{J}\Delta\Omega\approx 10^{-3} - 1$ for 
$\Delta\Omega=10^{-3}, 10^{-5}$. For dark matter clumps in the halo, 
Baltz {\it et. al.}~\cite{Peskin} estimate that a typical 
clump would have $\bar{J}\Delta\Omega\approx 0.4$ at 
$\Delta\Omega=1.5\times 10^{-4}$. (This estimate uses the data from a 
simulation by Taylor and Babul~\cite{TB}.) 
For a space-based telescope such as GLAST, the background flux for these 
targets can be estimated by a simple power-law extrapolation of the 
extragalactic gamma ray flux measured by EGRET~\cite{EGR}:
\beq
\frac{d\Phi}{dEd\Omega}\,=\,k\,\left(\frac{E}{{\rm 100~GeV}}\right)^{-2.1},
\eeq{EGRET_XG}
where $k=8.2\times 10^{-11}$ cm$^{-2}$s$^{-1}$sr$^{-1}$GeV$^{-1}$. 
Assuming a telescope with energy resolution $\delta E/E=0.1$, the
signal/background ratio is close to 1 for sources with 
$\bar{J}\Delta\Omega\sim 1$ in a model with $M\sim 100$ GeV. 
(The signal/background ratio decreases with increasing $M$ and/or
decreasing $\bar{J}$.) However, the number of $Z\gamma$ events expected 
at GLAST ($A=10^4$ cm$^2$) is well below 1 event/year, so that no discovery is 
possible. Of course, this pessimistic prediction could be proven wrong if 
dark matter turned out to be significantly stronger clumped at short 
scales than presently thought, resulting in larger values of $\bar{J}$.
Barring this possibility, a telescope with a larger effective area 
($A\gapproxeq 10^7$ cm$^2$) would be required to begin probing the heavy 
photon dark matter model in this channel.

The final annihilation process giving rise to monochromatic photon flux,
\B\B$\to h\gamma$, is strongly suppressed. It cannot occur via an $s$-channel 
Higgs exchange, because the process $h^\ast\to h\gamma$ is forbidden by the 
Ward identity of Quantum Electrodynamics. The proof is as follows:  
Let the photon and on-shell Higgs have momenta $k$ and $p$, respectively.  
The matrix element has the form
\beq
{\cal M}\left(h^\ast\to h\gamma\right)\equiv\varepsilon^\ast_\mu\left(k
\right){\hat{\cal M}}^\mu\,,
\eeqn
where ${\hat{\cal M}}^\mu$ can be decomposed as
\beq
\hat{\cal M}^\mu=A\left(k,p\right)\,k^\mu+B\left(k,p\right)\,p^\mu\,.
\eeqn
The Ward identity requires that $k_\mu\hat{\cal M}^\mu=0$.  For the 
first term this is true since $k^2=0$.  For the second term, $p\cdot k=0$ 
only if the initial Higgs is on-shell (when the process is kinematically 
forbidden); thus, $B\equiv0$.  However, by the transverse nature of the 
polarization vector the first term provides no contribution to the matrix 
element; therefore, ${\cal M}\left(h^\ast\to h\gamma\right)\equiv 0$.  
As a result, \B\B$\to h\gamma$ may only proceed via box diagrams with T-odd 
and T-even fermions in the loop, which receive no resonant enhancement.
Moreover, the \B$\tilde{Q}q$ and \B$\tilde{L}l$ couplings are of order 
$g^\prime/10 \approx 0.035$, further suppressing the cross section.

\section{Neutrino Fluxes from \B~Annihilations in the Sun and the Earth}

Neutrinos produced in annihilations of heavy photons collected in the 
gravitational wells of the Sun and the Earth provide another potentially 
observable indirect signature of Little Higgs dark matter. The 
procedure for evaluating the neutrino fluxes in a given model is well 
established; a thorough review (in the context of SUSY) is given 
in~\cite{bigneu}. Here we will follow this procedure to compute the 
neutrino fluxes expected in the LHT model.

The number of WIMPs $N$ collected in the Sun or the Earth obeys 
\beq
\dot{N}=C-A\,N^2,
\eeq{ndot}
where $C$ is the capture rate and $A$ is the annihilation rate per WIMP. 
Schematically,
the capture rate is given by
\beq
C \sim c \frac{\left< \sigma^{\rm el}_{\rm pb} \right>}{M^2_{\rm GeV}} 
\eeq{capture}
where $M_{\rm GeV}$ is the WIMP mass in units of GeV, the quantity 
$\left<\sigma^{\rm el}_{\rm pb}\right>$ is essentially the weighted average 
of the elastic WIMP-nucleus scattering cross sections (in pb) over the atomic 
composition of the Sun or the Earth, and $c$ is a coefficient determined by
the properties of the astronomical body in question: $c\sim 10^{30}$
s$^{-1}$ for the Sun and $c\sim 10^{20}$ s$^{-1}$ for the Earth. (See
Ref.~\cite{bigneu} for a more detailed discussion.) The only input from 
particle physics required to compute $C$ is the elastic scattering 
cross sections, which were computed in Ref.~\cite{BPNS} for the LHT model.
The annihilation rate per WIMP $A$ is schematically given by
\beq
A=\frac{\left<\sigma_{\rm an} u\right>}{V_{\rm eff}},
\eeq{annih}
where $\sigma_{\rm an}$ is the total WIMP annihilation cross section, 
the average is over the thermal distribution of the WIMPs captured in
the Sun or the Earth, and $V_{\rm eff}$ is the effective volume of the
``WIMP-sphere'' inside the astronomical body. (For details, see
Ref.~\cite{bigneu}.) 

Having computed $C$ and $A$, we solve Eq.~\leqn{ndot} to obtain the
total WIMP annihilation rate:
\beq
\Gamma_A=\frac{1}{2}~C~\tanh^2\left(t\,\sqrt{A\,C}\right)\,,
\eeq{anrt}
where $t\approx 4.5\cdot 10^9$ years is the age of the Solar System. The 
experimental technique best
suited to searching for high-energy neutrinos from WIMP annihilation relies on 
observing an upward-going muon created by a charged-current interaction of 
a muon neutrino in the rock below the detector. 
The rate of such muons per unit detector area is given by~\cite{bigneu}
\beq
\Gamma_{\rm detect} \,=\, (2.54\times 10^{-29}~{\rm m}^{-2}{\rm yr}^{-1})
\,\frac{\Gamma_A}{{\rm s}^{-1}}\,M_{\rm GeV}^2
\,\sum_i a_i b_i \sum_F B_F \left< Nz^2 \right>_{F,i}(M)\,,
\eeq{detect}
where $i$ are the possible neutrino types, $a_i$ and $b_i$ are (known)
coefficients describing the neutrino scattering and muon propagation in the 
rock, and $F$ are the possible final states of WIMP annihilation with 
branching fractions $B_F$. In the LHT model, the dominant annihilation
channels are $W^+W^-$ and $ZZ$. The quantity 
\beq
\left< Nz^2 \right>_{F,i}(M)\equiv \int_{E_{th}/M}^1 \frac{dN_{F,i}}{dz}(E_i,z)
z^2 dz 
\eeq{secmom}
is the second moment of the spectrum of neutrino type $i$ from final 
state $F$. Here $E_{th}$ is the threshold energy of the detector, and 
$dN_{F,i}/dz$ is the neutrino spectrum, normalized per single 
WIMP annihilation into the final state $F$. This spectrum is a convolution of 
the initial neutrino spectrum at the production point with the propagation 
effects (including neutrino oscillations and absorbtion) on the way to the
detector. In this analysis we used the neutrino spectra computed by 
Cirelli~{\it et.~al.}~\cite{Cirelli},\footnote{We thank Marco 
Cirelli for providing us with the updated version of the spectra.} 
and assumed
a detector with a threshold energy of 50 GeV, representative of the IceCube
experiment~\cite{ICECUBE}. Note however that lowering this threshold would 
not have a substantial effect on the rates, since the sub-threshold 
contribution to the rate scales as $(E_{\rm th}/M)^3$ and is at most of order 
10\% throughout the interesting parameter range in the LHT model.

\if
It includes the effects of
neutrino propagation in the Sun or the Earth. Analytic expressions for
the functions $\left< Nz^2 \right>_{F,i}(E_i)$ for the final states that
dominate WIMP annihilation in the LHT model ($F= W^+W^-, ZZ, t\bar{t}$)
are presented in Ref.~\cite{bigneu}. These expressions do not include the
detector threshold effects, and thus overestimate the rate. However, the
sub-threshold contribution to the rate scales as $(E_{\rm th}/M)^3$, where 
$E_{\rm th}$ is the threshold energy of a given detector. For example, for 
the ICECUBE detector~\cite{ICECUBE}, $E_{\rm th}\sim 50$ GeV, and the 
sub-threshold contribution is expected to be at most of order 10\% throughout 
the interesting parameter range in the LHT model.
\fi

\begin{figure}
\begin{center}
\includegraphics[width=10cm]{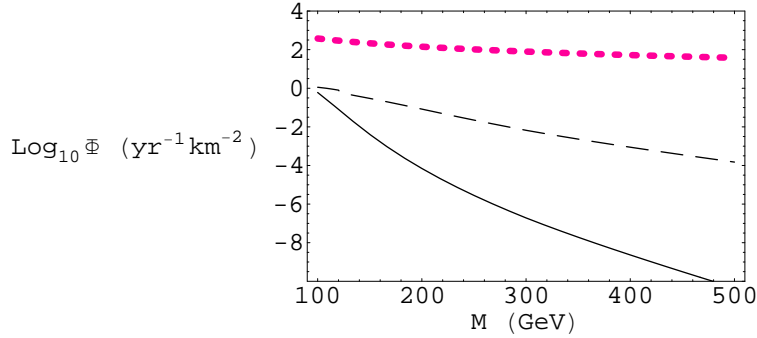}
\vskip-12mm
\caption{The rate of neutrino-induced upward-going muon events expected from 
the heavy photon annihilation in the Sun. Solid line: ``low'' 
pair-annihilation region (the flux in the ``high'' pair-annihilation 
region is similar); dashed line: coannihilation region with
$m_h=300$ GeV. Also shown is the expected sensitivity of the IceCube 
detector (red/dotted line).}
\label{fig:nuS}
\end{center}
\end{figure}

The rate of neutrino-induced upward-going muon events expected from the
heavy photon annihilation in the Sun is shown in Fig.~\ref{fig:nuS}.
The expected sensitivity of the IceCube detector is shown
for comparison. The maximum possible rate (achieved at the low end of the 
allowed LTP mass range, $M\approx 100$ GeV) is about 1 event/yr/km$^2$ in the
coannihilation region and about 0.5 events/yr/km$^2$ in pair-annihilation 
regions. Unfortunately these rates are well below the sensitivity of the
IceCube. The sensitivity would need to be improved by a factor of a few
hundred to a thousand before the fluxes predicted in the LHT models can
be probed. The rates of events due to heavy photon annihilation in the 
Earth are even smaller, of the order 
$10^{-5}$ events/yr/km$^2$ or below throughout the parameter space.

\section{Conclusions}

In this paper, we discussed two signatures of the heavy photon dark matter
predicted by the LHT theory. We computed the cross section of the process
$B_HB_H\to Z\gamma$, leading to a monochromatic gamma ray signature. We
also computed the flux of high-energy neutrinos from the annihilation of 
the heavy photons trapped in the Sun and the Earth. Unfortunately, the
near-term prospects for observing both signatures are rather poor.
the gamma ray signature could in principle be observed by GLAST, but only
if dark matter is very strongly clumped at short distance scales. 
In the neutrino case, the predicted flux is too small to be observed 
at the IceCube. 

This study complements two previous analyses of the discovery prospects for 
the LHT dark matter. While all predictions are subject to significant 
astrophysical uncertainties, it appears that the most promising search 
channels are the ``secondary'' gamma rays produced in hadronization and 
fragmentation of the primary WIMP annihilation products~\cite{BPNS}, and 
anomalous
high-energy positrons~\cite{Okada}. In these channels, the signal may be 
observed by the near-future instruments, GLAST in the case of gamma rays and
PAMELA and AMS-02 in the case of positrons. On the other hand, the LHT model 
predicts that no signal will be observed in the near-future direct detection
and high-energy neutrino searches, while the 
monochromatic gamma ray signal is very unlikely. By testing the predicted
pattern of signals, astroparticle experiments will provide an important
test of the LHT dark matter hypothesis, complementary to the more direct 
searches for the new particles predicted by the LHT model at the Large 
Hadron Collider.

{\it Acknowledgments ---} We are grateful to Dan Hooper for helpful
correspondence, and to Andrew Noble for useful discussions. We thank
Marco Cirelli for clarifying the results of Ref.~\cite{Cirelli} and
providing us with the updated version of the neutrino flux tables. This 
research is supported by the NSF grant PHY-0355005.

\end{document}